\documentstyle[13lomcon,cite,epsfig]{article}

\newcommand{\bee}{\begin{eqnarray}}
\newcommand{\eend}{\end{eqnarray}}

\newcommand{\Lar}{L_{\rm B}}
\begin{document}
\hspace{0.05cm}{\em  Talk given at the 13th Lomonosov Conference
on Elementary Particle Physics,

\hspace{0.05cm}
Moscow State University, 2007, August 23 - 29, Moscow}\\\\
\title{STRING-LIKE ELECTROSTATIC INTERACTION\\ FROM
QED WITH INFINITE MAGNETIC FIELD}

\author{A.E. Shabad \footnote{e-mail: shabad@lpi.ru}}

\address{P.N. Lebedev Physics Institute, Leninsky prospect 53, Moscow 119991,
Russia}

\author{V.V. Usov \footnote{e-mail: fnusov@wicc.weizman.ac.il}}

\address{Center for Astrophysics, Weizmann Institute
of Science, Rehovot 76100, Israel}

\maketitle
 \vspace{0.25cm}
 \abstracts{ In the limit of infinite external magnetic field $\bf B$ the static field of
   an electric charge is squeezed into a string parallel to $\bf B$.
   Near the charge the  potential grows like $|x_3|(\ln |x_3| + const)$ with
   the coordinate $x_3$ along the
   string. The energy of the string breaking is finite and
  very close to the effective photon mass.}

It is known that the attraction force between two colored charges,
when calculated using the Wilson loop method on a lattice, is
concentrated in a string with its width being of the order of the
lattice spacing $l$. The string potential consists of three additive
components: $(i)$ the Coulomb term $1/R$ that dominates at small
distance $R$ from the charge, $(ii$) a constant term, corresponding
to the infinite mass renormalization (recall that $l$ is the
ultraviolet cutoff parameter in the lattice theory), turns to
infinity as $1/l$, when  $l$ is taken close to zero,  $(iii$) the
term linearly growing with $R$ corresponding to a constant string
tension and providing the confinement according to the Wilson area
criterion (see, $e.g.$, \cite{kondo}).

We show that in quantum electrodynamics with external magnetic field
$\bf B$ the Coulomb potential of a point-like static charge, when
corrected by the vacuum polarization, acquires a similar string-like
form in the infinite-magnetic-field limit, the electron Larmour
radius $L_{\rm B}=(\sqrt{eB})^{-1}=(1/m\sqrt{b})\rightarrow 0$
playing, in a way, the same r\^{o}le as the spacing $l$ does in the
lattice theory. Here $b$ stands for the magnetic field, $b=B/B_0,$
measured in the units of $B_0=m^2/e=4.4\times 10^{13}$G, $e$ and $m$
are electron charge and mass, resp.

 Electric potential $A_0$ of a
point charge $q$ when calculated as a sum of chains of
electron-positron loops in a magnetic field $B$, so strong that the
Larmour radius is much smaller than the electron Compton length,
$\Lar \ll m^{-1}$ (this implies $B\gg B_0$), is represented as a sum
\bee\label{twoparts}A_0({\bf x})= A_{\rm s.r.}({\bf x})+A_{\rm
l.r.}({\bf x}).\eend The analytic expressions for these functions
are given in \cite{SU1, SU2}. The argument  $\bf x$ here is the
radius-vector with its origin in the charge, its components across
and along the magnetic field being ${\bf x}= ({\bf x}_\perp, x_3)$.

The function $A_{\rm s.r.}({\bf x})$ has the exact scaling
property\bee\label{scaling}\hspace{-1.5cm} A_{\rm s.r}({\bf x})=
\frac{\widetilde{A}(\widetilde{\bf x})}{L_{\rm B}},\eend where the
dimensionless function $\widetilde{A}$ contains the magnetic field
through its argument $\widetilde{\bf x}={\bf x}\Lar^{-1}$ only . The
function $A_{\rm s.r.}({\bf x})$ is short-range: for distances from
the charge, large in the Larmour scale, $|x_3|\gg \Lar$, or
$|x_\perp|\gg \Lar,$ it  reduces to the Yukawa law\bee\label{yuk}
A_{\rm s.r.}({\bf x})\simeq \frac {q}{4\pi L_{\rm B}}\frac
{\exp\{-\left(\frac{2\alpha}\pi\right)^{\frac 1{2}}
{\sqrt{\widetilde{x}_\perp^2+\widetilde{x}_3^2}}
\}}{\sqrt{\widetilde{x}_\perp^2+\widetilde{x}_3^2}}=\frac
q{4\pi}\frac{\exp\{-\left(2\alpha b/\pi\right)^{\frac 1{2}}m|{\bf
x}|\}} {|{\bf x}|},\eend  where $\alpha=e^2/4\pi=1/137$. This
equation reflects the Debye screening of the charge by the polarized
vacuum. The effective photon mass (inverse Debye radius) in
(\ref{yuk}) $M= (2\alpha/\pi)^{1/2}\;L_{\rm B}^{-1}$ tends to
infinity together with the magnetic field. The "photon mass
$\widetilde{M}\;$" in the dimensionless function
$\widetilde{A}(\widetilde{\bf x})$ is finite, ${\widetilde{M}}=
(2\alpha/\pi)^{1/2}$, and corresponds to the topological photon mass
in the two-dimensional massless electrodynamics of Schwinger
\cite{schwinger}. Anisotropic corrections to (\ref{yuk}) were
pointed in \cite{tehran}.

 The second
term in (\ref{twoparts}) is long-range: it slowly decreases at the
distances of the order of the Compton length $m^{-1}$ and larger,
following the anisotropic Coulomb law \bee\label{llargex}A_{\rm
l.r.}(x_3,x_\perp)\simeq \frac 1{4\pi}\frac
{q}{\sqrt{(x^\prime_\perp)^2+x_3^2}},\qquad
x^\prime_\perp=x_\perp\beta,~~~~\beta=\left(1+\frac{\alpha
b}{3\pi}\right)^{\frac 1{2}}.\eend It represents
 the whole potential $A_0({\bf x})$ (\ref{twoparts}) there, since
$A_{\rm s.r.}({\bf x})$ is already negligible at such distances.
This potential decreases with the growth of perpendicular distance
$x_\perp$ from the charge faster than along the field. The
equipotential surface is an ellipsoid $(x^\prime_\perp)^2+x_3^2=\;
$const, which contracts to a string in the limit $b=\infty$. The
family of electric lines of force, parameterized by the angle
$\phi,$  is for each value of the magnetic field given as
$x_3=(x_\perp)^{\beta^{-2}}\tan\phi$ (see Fig. 1). All of them
gather together inside a string passing through the charge and
directed along the external magnetic field.
\begin{figure}[t]
\centering
\includegraphics[scale=1.0]{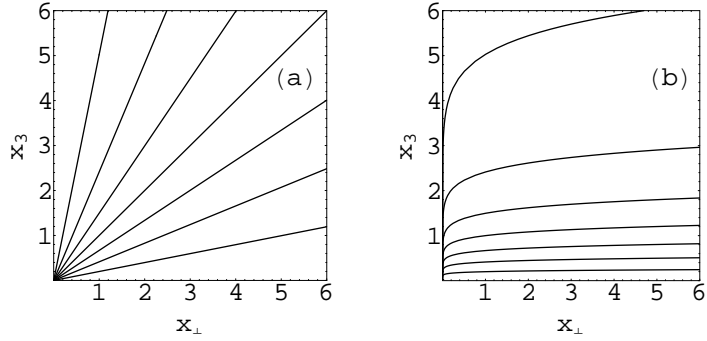}
\caption{\hspace{0.5cm}Lines of force of electric field coming from
a point charge.\qquad \qquad \qquad \qquad \qquad (a) No magnetic
field, $B$=0, (b) $B=10^4B_0$}
\end{figure}

The short-range part of the potential has the following asymptotic
expansion near the point $x_3=x_\perp=0,$ where the charge is
located,
 \bee\label{expand}A_{\rm s.r.}({\bf x})\simeq
\frac{q}{4\pi}\left(\frac 1 {|\bf x|}-2mC_{\rm
s.r.}+o(x_3^2)+o(x_\perp^2)\right).\eend  For large magnetic
fields $b\gg 2\pi/\alpha \sim 10^3$ we have $2mC_{\rm s.r.}\simeq
0.9595\;\Lar^{-1}\sqrt{2\alpha/\pi}$. It is infinite for $\Lar=0$
or $b=\infty$, as stated at the beginning. The corresponding
constant in the expansion of $\widetilde{A}(\widetilde{\bf x})$ is
finite: $2\widetilde{C}=0.9595\;\sqrt{2\alpha/\pi}$. The constant
0.9595 is calculated using the experimental value of $\alpha$. For
$\alpha=0$ it turns into unity. The growth of $C_{\rm s.r.}$ with
the magnetic field following the square root law provides the
narrowing of the potential and, in the end, the finiteness of the
ground state energy of a hydrogen-like atom in infinite magnetic
field (see \cite{SU1, SU2} and also the discussion \cite{wang}).
In the limit $b=\infty$ the short-range part of the potential
becomes the Dirac $\delta$-function: \bee\label{deltaxperp}
\left.A_{\rm s.r.}(0,x)\right|_{b=\infty}=2.178~\frac
q{2\pi}\delta(x).\eend The behavior of the long-range part $A_{\rm
l.r.}(x_3,x_\perp=0)$ at the string $x_\perp=0$ is shown in Fig.2.
(See \cite{SU2} for analytical equations). The limiting form of
this function at $\Lar =0$ is finite. It decreases in agreement
with Eq. (\ref{llargex}) at large distances, and has the following
behavior near the charge, $|x_3|\ll m^{-1}$,
\bee\label{confinement}\left.A_{\rm l.r.}(x_3,0
)\right|_{b=\infty}\simeq\left.A_{\rm l.r.}(0,0
)\right|_{b=\infty}\nonumber\\+\frac{qm}{4\pi}\left(1-{\alpha\over\pi}
f(\alpha)\right)2m|x_3|\left[\ln (2m|x_3|)-{1\over 2}\ln
2+\gamma-1\right],\quad \eend where $\gamma=0.577$ is the Euler
constant and the coefficient $f$ depends on the fine structure
constant,  $f\left(\alpha={1/137.036}\right)=4.533$. It is
nonanalytic in $\alpha$
:\\$\left.f(\alpha)\right|_{\alpha\rightarrow 0}\simeq
-\ln\alpha$. We see that the growth is not linear. It provides
"confinement" within the Compton distances, where the
approximation (\ref{confinement}) is valid. The first constant
term in (\ref{confinement}) is defined by an integral depending on
the fine structure constant. If calculated with its experimental
value, $\alpha=1/137.036,$ it makes $\left.A_{\rm l.r.}(0,0
)\right|_{b=\infty}=1.4152(qm/2\pi)$. The numerical coefficient
here is rather close to $\sqrt{2}=1.4142$. Bearing in mind that
$\left.A_{\rm l.r.}(\infty,0 )\right|_{b=\infty}=0$ we note that
$\left.A_{\rm l.r.}(0,0 )\right|_{b=\infty}$ is the increment of
the long-range part of the potential along the string between the
point where the charge is located and the infinitely remote point,
i.e. is equal to the work needed for removing a  unit test charge
to infinity. It may be referred to as the energy of the string
breaking. It is remarkable that this quantity is finite. If we set
$q=e$, we find that the  energy density of breaking of a string
associated with the electron is with surprising accuracy equal to
the dimensionless photon mass : $\left.A_{\rm l.r.}(0,0
)\right|_{b=\infty}/m=1.0007~\widetilde{M}$. The coincidence would
be exact for the value of the fine-structure constant equal to
$1/121$. By postulating this coincidence as a a physical principle
we may obtain an approach for calculating $\alpha$, to which end
an approximation, better than one-loop approximation referred to
here, would be needed.
\begin{figure}[t]
\centering
\includegraphics[scale=0.7]{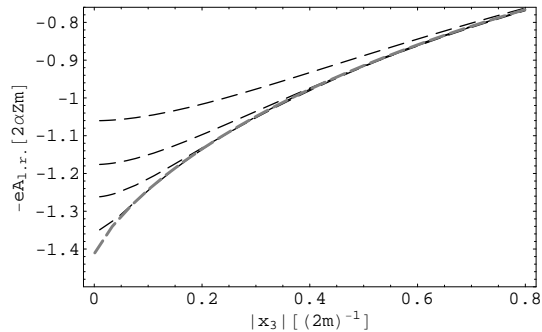}
\caption{Energy $-eA_{\rm l.r.}(x_3,0)$ of the electron in the
long-range part  of the potential of the point charge $q=eZ$ for
$b=10^6,\;
    10^5$, $3\times 10^4$, $10^4$ (dashed lines from bottom to top).
    The dashed thick line
    corresponds to the limit $b=\infty$ and represents the string
    potential.}
\end{figure}

Formation of a string in QED discussed in the present paper is not
completely unexpected, since it was noted \cite{kondo} that the
Abelian theory does have a topologically nontrivial two-dimensional
sector (described by a nonlinear $\sigma$-model).
\section*{Acknowledgments}
 Supported by Program
No. LSS-4401.2006.2, RFBR Project No. 05-02-17217, and by the Israel
Science Foundation of IASH. One of the authors (A.E.S.) expresses
his gratitude to Professor D.B.Melrose for hospitality at the
University of Sydney and valuable discussions.


\section*{References}
\begin{thebibliography}{20}
\bibitem{kondo} Kei-Ichi Kondo, {\it Phys.Rev.} {\bf D} 58, 085013
(1998).
\bibitem{SU1} A.E. Shabad and V.V. Usov, {\it Phys.Rev.Lett.} {\bf 98},
180403 (2007).
\bibitem{SU2} A.E. Shabad and V.V. Usov, arXiv: 0707.3475
[astro-ph].
\bibitem{schwinger} J. Schwinger, {\em Phys. Rev.} {\bf 128},
 2425 (1962).
\bibitem{tehran} N. Sadooghi and A. Sodeiri Jalili, {\it Phys.Rev.}
{\bf D} 76, 065013 (2007).
\bibitem{wang} S.-Y. Wang, {\em Phys.Rev.Lett.} {\bf 99}, 228901
(2007); A.E. Shabad and V.V. Usov, {\it Phys.Rev.Lett.} {\bf 99},
 228902 (2007).
\end{thebibliography}
\end{document}